\documentclass[preprintnumbers,10pt,nofootinbib]{revtex4}

\usepackage{amsmath,latexsym,amssymb,amsfonts}
\usepackage[pdftex]{color,graphicx}
\begin{document}

%\preprint{}

\title{\textbf{Suppression of long-wavelength CMB spectrum from the Hartle-Hawking wave function in Starobinsky-type inflation model}}
\author{
\textsc{Pisin Chen$^{a,b,c,d}$}\footnote{{\tt pisinchen{}@{}phys.ntu.edu.tw}},
\textsc{Hsiao-Heng Yeh$^{a,b}$}\footnote{{\tt b01202056{}@{}ntu.edu.tw}} 
and
\textsc{Dong-han Yeom$^{e,f,g}$}\footnote{{\tt innocent.yeom{}@{}gmail.com}}
}

\affiliation{
$^{a}$\small{Leung Center for Cosmology and Particle Astrophysics, National Taiwan University, Taipei 10617, Taiwan}\\
$^{b}$\small{Department of Physics and Center for Theoretical Sciences, National Taiwan University, Taipei 10617, Taiwan}\\
$^{c}$\small{Graduate Institute of Astrophysics, National Taiwan University, Taipei 10617, Taiwan}\\
$^{d}$\small{Kavli Institute for Particle Astrophysics and Cosmology,
SLAC National Accelerator Laboratory, Stanford University, Stanford, California 94305, USA}\\
$^{e}$\small{Asia Pacific Center for Theoretical Physics, Pohang 37673, Republic of Korea}\\
$^{f}$\small{Department of Physics, POSTECH, Pohang 37673, Republic of Korea}\\
$^{g}$\small{Department of Physics Education, Pusan National University, Busan 46241, Republic of Korea}
}

\begin{abstract}
The lack of correlations on the large scale cosmic microwave background (CMB) anisotropy provides a potential window to probe beyond the standard inflationary scenario. In this paper, we investigate the primordial power spectrum based on the Hartle-Hawking (HH) no-boundary proposal for a homogeneous, isotropic, and spatially-closed universe that leads to a Starobinsky-type inflation after the classicalization. While we found that there is no suppression at large scales in the standard $R+R^2$ theory, we also found that it is possible to sufficiently suppress the large-scale power spectrum if a pre-inflation stage is introduced to the Starobinsky-type model. We calculate the $C^{TT}_{\ell}$ correlation function and show that our proposal gives a better fit to the Planck CMB data. This suggests that our universe might have begun with a compact HH state with a small positive curvature.
\end{abstract}

\maketitle

\newpage

\tableofcontents

%\newpage

%%%%%%%%%%%%%%%%%%%%%%%%%%%%%%%%%%%%%%%%%%%%%%%%%%%%%%%
\section{Introduction}

In spite of the great success of modern inflationary cosmology \cite{Guth1981}, we barely understand physics before inflation and its quantum mechanical origin. Unfortunately, this is not only because of the poor understanding of the physical nature about quantum theory of gravity but also of the limited observational data within our causal horizon. The current largest distant structure that we can see is from the cosmic microwave background \cite{Planck2016XX}, which is a fingerprint of the cosmic DNA at the time of recombination and provides a rich amount of information. To trace further back in time to peek into the beginning of the universe, the observations of the primordial gravitational waves through the CMB polarizations may provide a precious window.

The 2018 Planck data shows that the lack of correlation at large scale CMB anisotropy remains persistent, i.e., the CMB correlation power spectrum below $\ell=30$ exhibits deviation from the standard Lambda-CDM ansatz \cite{Akrami:2018odb}. While such deviation is not yet statistically significant, and the mainstream conviction is that this is caused by the cosmic variance, we argue that the possibility of having this low-mode anomaly induced by an underlying physics should not be brushed aside prematurely. In the standard slow-roll inflation scenario, the power spectrum is nearly scale-invariant because the spacetime is quasi-de Sitter. By calculating the vacuum polarization of the inflaton field in terms of the canonical quantization in the Bunch-Davies (BD) vacuum \cite{Bunch1978}, one can expand the spectrum into Fourier modes such that each mode will be frozen once exit the Hubble horizon. In quasi-de Sitter space, the exponential expansion will lead to an almost constant rate of exit of modes, which is why the spectrum is nearly scale-invariant. However, if the universe is not perfectly flat, then large scale modes, whose sizes are comparable to the Hubble scale, may not satisfy the exact scale-invariance \cite{Chen:2017aes,White:2014aua}. As a result, the large-scale modes may in principle exhibit some deviation from that predicted by the standard approach \cite{Chen:2015gla}. 

In our previous analysis \cite{Chen:2017aes}, we invoked a minimally coupled scalar field with a quadratic potential as a working inflationary model \cite{Linde1983}. In order to assign quantum states so as to carry out their perturbations, we assume that the quantum state is given by Hartle-Hawking (HH) wave function \cite{Hartle1983}, which is one of the proposals to the boundary condition of the Wheeler-DeWitt equation \cite{DeWitt1967}. Applying the steepest-decent approximation, we approximated the wave function of the universe as a sum over compact instanton solutions to the $\phi^2$ inflaton model. Based on this method, we demonstrated that the large scale power spectrum could be suppressed if the mass of the inflaton field is sufficiently large comparable to the Hubble scale. 

In this paper, we extend our previous result to the Starobinsky model \cite{Starobinsky1980}, which is favored by the Planck data \cite{Akrami:2018odb}. As we will show, it turns out that the large scale spectrum would be enhanced rather than suppressed in the original Starobinsky model. We found, however, that the suppression of large-scale modes is attainable if one generalizes the Starobinsky model to include a pre-inflation stage in the inflaton potential.

We emphasize that our approach is consistent with the Bunch-Davies vacuum without any artificial tunings. This is in contrast with some other approaches whose vacuum is not invariant (see, for example, \cite{Chen:2015gla}). We evaluate the quantum gravitational perturbations based on the Euclidean path integral approach, where the amplitudes of the perturbations are related to the Hawking temperature. If the universe was a bona fide de Sitter space, then the inflation would be eternal and the Hawking temperature, which is proportional to the Hubble parameter, must be a constant in time. In reality, however, the cosmic inflation lasted for only a finite time and the Hubble parameter during inflation must decrease slowly and monotonically. As we will show in this paper, if the universe began as a Hartle-Hawking state, then there should be a region at the turning point of the Euclidean-Lorentzian junction where the effective Hawking temperature vanishes. After this moment, the Hawking temperature, and therefore the corresponding Hubble parameter, will increase to approach the canonical Hawking temperature in the static de Sitter limit. Such a transition would therefore result in the apparent suppression of the long wavelength modes in the CMB power spectrum without changing the main body of it based on the standard slow-roll inflation. We think that this is a natural way to explain how the universe came into being. 

This paper is organized as follows. In Sec.~\ref{sec:fro}, we introduce the techniques to derive the power spectrum based on  HH proposal. In Sec.~\ref{sec:citeref}, we calculate the gauge invariant power spectrum with the Starobinsky potential and our generalized version numerically, and then connect it to the CMB observation. Finally, in Sec.~\ref{sec:conclu}, we summarize our results and discuss about possible future directions. In this paper, we use the Planck unit ($\hbar=c=G=1$) with the spacetime signatures $(-,+,+,+)$.

%%%%%%%%%%%%%%%%%%%%%%%%%%%%%%%%%%%%%%%%%%%%%%%%%%%%%%%
\section{\label{sec:fro}Power spectrum from the no-boundary wave function}

\subsection{The Hartle-Hawking no-boundary proposal}

In order to assign quantum states based on quantum gravitational principles, we invoke Hartle-Hawking no-boundary wave function. According to Hartle and Hawking \cite{Hartle1983}, the wave function of the universe $\Psi[h_{ij},\Phi]$ for the three-geometry with metric $h_{ij} = g_{\mu\nu}|_{\partial \mathcal{M}}$ and the field configuration $\Phi = \tilde{\phi}|_{\partial \mathcal{M}}$ with a given compact manifold $\mathcal{M}$ is formally given by the Lorentzian path integral
\begin{equation}
	\Psi[h_{ij},\Phi]=\int_{\mathcal{M}} \mathcal{D}g_{\mu\nu} \mathcal{D}\tilde{\phi}\;\; e^{iS[g_{\mu\nu},\tilde{\phi}]},
\end{equation}
where the action for a minimally coupled scalar field $\tilde{\phi}$ and the Einstein gravity with the metric $g_{\mu\nu}$:
\begin{equation}
	S = \int d^4x \sqrt{-g} \left[ \frac{R}{16\pi} -\frac{1}{2}\partial_\mu\tilde{\phi} \partial^\mu\tilde{\phi} - \tilde{V}(\tilde{\phi}) \right].
\end{equation}
In order to describe the ground state of the wave function, one can Wick-rotate the time such that $t=i\tau$ to work in the Euclidean space with the corresponding action (hence, $iS=S_{\mathrm{E}}$). Then the wave function becomes
\begin{equation}
	\Psi[h_{ij},\Phi]=\int_{\mathcal{M}} \mathcal{D}g_{\mu\nu} \mathcal{D}\tilde{\phi}\;\; e^{-S_{\mathrm{E}}[g_{\mu\nu},\tilde{\phi}]},
\end{equation}
where we integrate all compact Euclidean four-manifolds $\mathcal{M}$ that have $[h_{ij},\Phi]$ as their boundary.

This wave function can be further approximated by introducing the minisuperspace approximation and the steepest-descent approximation. We focus on the homogeneous, isotropic, and spatially closed universe. Then, the Euclidean metric can be written as
\begin{equation} \label{metric}
	ds^2=\sigma^2 \left[ \left(N^{2} - N_{i}N^{i} \right)d\lambda^2 + 2 N_{i} dx^{i} d\lambda + a^{2}(\lambda)\gamma_{ij} dx^{i}dx^{j} \right] = \sigma^2 \left[ N^{2}d\lambda^2+a^2(\lambda) d\Omega_3^2 \right],
\end{equation}
where $\sigma$ is a normalization constant, $N$ is the lapse function and $N_{i}$ is the shift function (we choose $N = \bar{N}$ and $N_{i} = 0$ as the background solution), $a$ is the scale factor, and $d\Omega_{3}^{2}$ is the infinitesimal solid angle in the three-sphere. Let us simplify the notations by redefining
\begin{eqnarray}
\phi &=& \sqrt{\frac{4\pi}{3}} \tilde{\phi},\\
V &=& \frac{8\pi \sigma^{2}}{3} \tilde{V}(\tilde{\phi}).
\end{eqnarray}
We can then recast the path integral as 
\begin{equation}
	\Psi=\int \mathcal{D}N \mathcal{D}a \mathcal{D}\phi\;\; e^{-S_{\mathrm{E}}[N,a,\phi]},
\end{equation}
where
\begin{equation}
	S_{\mathrm{E}}[N,a,\phi]=\frac{3\pi\sigma^2}{4}\int d\lambda \; N\left[-a \left(\frac{\frac{da}{d\lambda}}{N}\right)^2-a+a^3\left(\left(\frac{\frac{d\phi}{d\lambda}}{N}\right)^2 - V(\phi)\right)\right].
\end{equation}
Then the equations of motion can be obtained as follows:
\begin{eqnarray} \label{bgeom1}
	{a'}^2-1+a^2 \left(-{\phi'}^2+ V\right) &=& 0,\\
\label{bgeom2}
	a''+2a{\phi'}^2+a V &=& 0,\\
 \label{bgeom3}
	{\phi''}+3 \frac{a'}{a}{\phi'}-\frac{1}{2}\frac{d V}{d\phi} &=& 0,
\end{eqnarray}
where the prime ($'$) denotes a derivation with respect to $\tau$, which is defined by $d\tau=Nd\lambda$. One can solve an on-shell solution by imposing the compactness ($a(0) = 0$) and regularity (${a'}(0)=1$ and ${\phi'}(0) = 0$) conditions to the Euclidean manifold. The solution will be given in Sec.~\ref{sec:citeref}.

%%%%%%%%%%%%%%%%%%%%%%%%%%%%%%%%%%%%%%%%%%%%%%%%%%%%%%%
\subsection{Harmonic expansion in a closed universe}

Following Halliwell and Hawking \cite{Halliwell1985}, one can expand the perturbations on a spacelike hypersurface with the $S^3$ topology into spherical harmonics. We first separate $\gamma_{ij}$ into the background part and the perturbation part:
\begin{equation}
	\gamma_{ij}=\bar{\gamma}_{ij}+\epsilon_{ij},
\end{equation}
where $\bar{\gamma}_{ij}$ denotes the background three-sphere and $\epsilon_{ij}$ denotes the perturbation of the three-sphere. Here, the perturbation is expanded by
\begin{eqnarray}
	\epsilon_{i j} &=& \sum_{n,\ell,m} \left[ \sqrt{6}q_{n\ell m}\frac{1}{3}\bar{\gamma}_{ij}Q_{n\ell m}+\sqrt{6}b_{n\ell m}(P_{ij})_{n\ell m} +\sqrt{2} c^o_{n\ell m}(S^o_{ij})_{n\ell m}+\sqrt{2} c^e_{n\ell m}(S^e_{ij})_{n\ell m} \right. \nonumber \\
	&& \left. + 2d^o_{n\ell m}(G^o_{ij})_{n\ell m}+2d^e_{n\ell m}(G^e_{ij})_{n\ell m} \right],
\end{eqnarray}
where $n$, $\ell$, and $m$ are the spherical coordinate indices. In addition, the lapse and shift are expanded by
\begin{eqnarray}
	N &=& \bar{N} \left[ 1+ \sum_{n,\ell,m} \frac{1}{\sqrt{6}}g_{n\ell m}Q_{n\ell m} \right],\\
	N_i &=& a \left[ \sum_{n,\ell,m} \frac{1}{\sqrt{6}}k_{n\ell m}(P_i)_{n\ell m}+ \sum_{n,\ell,m} \sqrt{2}j_{n\ell m}(S_i)_{n\ell m} \right].
\end{eqnarray}
Here, $q$, $b$, $c^o$, $c^e$, $d^o$, $d^e$, $g$, $k$, and $j$ are time dependent coefficients, while $Q$, $P_{i j}$, $S^o_{i j}$, $S^e_{i j}$, $G^o_{i j}$, $G^e_{i j}$, $P_{i}$, and $S_{i}$ are space dependent basis.
Note that
\begin{eqnarray}
	P_{ij} &=& \frac{1}{n^2-1}\nabla_i\nabla_j Q+\frac{1}{3}\bar{\gamma}_{ij}Q,\\
	P_i &=& \frac{1}{n^2-1}\nabla_iQ,
\end{eqnarray}
where the covariant derivatives are with respect to $\bar{\gamma}_{ij}$.

%Note that scalar, vector, and tensor modes are decoupled at the second order. (?) Hence, we only keep track of $q$ and $b$ terms since they correspond to the scalar perturbation, while $c^{o}$, $c^{e}$, $d^{o}$, and $d^{e}$ terms correspond to vector and tensor modes.

Since we can only keep track of the scalar perturbation, one can neglect $c^{o}$, $c^{e}$, $d^{o}$, $d^{e}$, and $j$ terms. Moreover, one can further choose a gauge $q = b = 0$. Then we need to match the perturbations with the scalar field:
\begin{equation}
	\phi = \bar{\phi}+\sqrt{2}\pi \sum_{n,\ell,m} f_{n\ell m}Q_{n\ell m},
\end{equation}
where $\bar{\phi}$ is the background solution of $\phi$. Finally we reduce the relation between $f$, $k$, and $g$ coefficients. Especially, with the slow-roll limit and the linear approximation, the equation for $f$ modes are written by \cite{Halliwell1990}
%\begin{equation} \label{eom}
%	\ddot{f}_{n\ell m}+3H\dot{f}_{n\ell m}+ \left( \left( \frac{\partial V}{\partial\phi} \right)^2+\frac{n^2-1}{a^2} \right) f_{nlm}=-2 \left( \frac{\partial V}{\partial\phi} \right)^2\phi g_{n\ell m}+\dot{\phi}\dot{g}_{n\ell m}-\frac{\dot{\phi}}{Na}k_{n\ell m},
%\end{equation}
\begin{equation} \label{eom}
	\ddot{f}_{n\ell m}+3 \frac{\dot{a}}{a} \dot{f}_{n\ell m}+ \left( \frac{\partial^2 V}{\partial\phi^2}+\frac{n^2-1}{a^2} \right) f_{n\ell m} = 0,
\end{equation}
where the dot ($\dot{~}$) denotes a derivation with respect to the Lorentzian time $t=i\tau$.

\subsection{Calculating cosmological observables}

Now one can calculate the power spectrum of each mode. Using minisuperspace approximation, we expand the action $S_{\mathrm{E}} = \bar{S} + \sum_{n\ell m}S_{n\ell m}$, where $\bar{S}$ is the background action and $S_{n\ell m}$ is the perturbed action of the mode $(n,\ell,m)$. The action of each mode is that
\begin{equation}
	S_{n\ell m} \simeq \frac{a^3 \sigma^{2}}{2} \left( f_{n\ell m}\frac{df_{n\ell m}}{dt}-\frac{d\phi}{dt}g_{n\ell m}f_{n\ell m} \right).
\end{equation}
By using the trick of Laflamme \cite{Laflamme1987}, one can further specify the Euclidean vacuum condition for the mode $(n,\ell,m)$:
\begin{equation}
\Psi_{n\ell m} \left[f_{n\ell m} \right] \simeq B_{n\ell m} \exp \left[ - \frac{a^{3} \sigma^{2}}{2} \frac{\dot{\bar{f}}_{n\ell m}}{\bar{f}_{n\ell m}} f_{n\ell m}^{2} \right],
\end{equation}
where $B_{n\ell m}$ is a normalization constant, $\bar{f}_{n\ell m}$ is the solution evaluated at the horizon crossing time ($aH \simeq n$) and $f_{n\ell m}$ is regarded as a variable of the wave function $\Psi_{n\ell m}$. By using this wave function, we can evaluate the expectation value of $f_{n\ell m}$:
\begin{eqnarray}
	\langle f_{n\ell m}^2\rangle &=& \int_{-\infty}^{\infty}df_{n\ell m}f_{n\ell m}^2 \left| B_{n\ell m} \exp \left[ - \frac{a^{3} \sigma^{2}}{2} \frac{\dot{\bar{f}}_{n\ell m}}{\bar{f}_{n\ell m}} f_{n\ell m}^{2} \right] \right|^{2}\\
	&=& \frac{\bar{f}_{n\ell m}}{2a^3 \sigma^{2} \dot{\bar{f}}_{n\ell m}}.
\end{eqnarray}
Finally, the matter perturbation over the physical space is given by
\begin{eqnarray} 
	\langle\delta\tilde{\phi}^2\rangle &=& \frac{1}{2\pi^2}\int d\chi d\theta d\varphi \; \sin^2\chi\sin\theta\times\frac{3\pi}{2}\sum_{n\ell m}\sum_{n'\ell' m'}\langle f_{n\ell m}f_{n'\ell' m'}\rangle Q_{n\ell m}Q_{n'\ell' m'} \\
	&=& \frac{3}{4\pi}\sum_{n\ell m}\langle f_{n\ell m}^2\rangle \\
	&=& \sum_n\frac{3n^2}{4\pi} \langle f_n^2\rangle. \\
	&=& \sum_n\frac{nP(n)}{n^2-1}.
\end{eqnarray}
Where the dimensionless power spectrum is defined by \cite{Chen:2017aes}
\begin{equation}
	P(n) = \frac{3n(n^2-1)}{8\pi\sigma^2 a^3} \frac{\bar{f}_n}{\dot{\bar{f}}_n}.
\end{equation}

%%%%%%%%%%%%%%%%%%%%%%%%%%%%%%%%%%%%%%%%%%%%%%%%%%%%%%%
\section{\label{sec:citeref}Numerical solution and connection with CMB observations}

\subsection{Solving equations}

%Before introducing the numerical method, we want to emphasize that we are doing path integral in terms of "perturbation", leaving the background quantity fixed by the boundary condition, e.g. different types of instanton. This is different from what Hawking's proposal \cite{} since we are interesting in whether large scale mode, those curvature radius is comparable to Hubble radius, will develop some features in the CMB spectrum and perhaps the origin for low multiple anomaly. We will show that for small scale mode, the power spectrum reduce to standard BD vacuum vacuum which is nearly scale-invariant. For fix background, the extreme path $\{a_{ext}, \phi_{ext}\}$ is calculated by the equation of motion \eqref{bgeom2} and \eqref{bgeom3}. By choosing standard no boundary initial condition \footnote{All the background quantity on the Euclidean spacetime will be denoted with $\hat{}$ together with Euclidean time coordinate $d\tau=Nd\lambda$, leave the quantity on Lorentzian spacetime normal with physical time $t$ .}

In order to calculate the power spectrum, we need to first solve the background solution. As we invoke the compactness and the regularity conditions at the south pole of the $S^3$ Euclidean space, we impose
\begin{eqnarray}
	a(\tau=0) &=& 0, \\
	\dot{a}(\tau=0) &=& 1, \\
	\dot{\bar{\phi}}(\tau=0) &=& 0,
\end{eqnarray}
where $\bar{\phi}(0)$ is a free parameter. In the slow-roll limit, we approximate background solution as
\begin{equation}
	\hat{a}(\tau)=\frac{1}{H_0}\sin{H_0\tau},
\end{equation}
where $H_0$ is the integration constant which will be identified as the Hubble parameter at the de Sitter expansion stage. After Wick rotating to the Lorentzian signatures at the equator of the $S^3$ sphere, i.e.,
\begin{equation}
	\tau=\frac{\pi}{2H_0}+it,
\end{equation}
the matching conditions are
\begin{eqnarray}
	a(t=0)&=&\frac{1}{H_0}, \\
	\dot{a}(t=0)&=&0, \\
	\dot{\bar{\phi}}(t=0)&=&0.
\end{eqnarray}
Where $\bar{\phi}(t=0)$ is a free parameter, which controls the number of $e$-foldings. %For $N\approx 60$, we require that $\phi(t=0)\approx 1.2$.

%By solving the equation of motion, the background evolution corresponds to a standard slow-roll inflationary de-Sitter universe.
%Then we will do perturbation calculation in terms of this background. To fix the value of $H_0$, we consider the Hamiltonian constraint in Lorentzian space,
%\begin{equation}
%	\frac{\dot{a}^2}{a^2}=H^2=\frac{8\pi\sigma^2}{3}V(\phi)-\frac{1}{a^2}+\frac{4\pi}{3}\dot{\phi}^2.
%\end{equation}
With slow-roll approximation, we can assume that the potential is nearly a constant $V(\phi)\approx V_0$, then the Hubble parameter is approximately
\begin{equation}
	H \simeq \sqrt{\frac{8\pi\sigma^2}{3}V_0}.
\end{equation}
Hence, we can choose the metric normalization constant
\begin{equation}\label{eq:sigmaConvention} 
	\sigma^2=\frac{1}{V_0}.
\end{equation}
As a result, during the exponential growth, $H \simeq H_0 \simeq \sqrt{8 \pi / 3}$. Using Hubble constant $H$, we can numerically solve equations of motion for $f_{n}$.

%To evaluate the power spectrum, we trace through the background geometry with a Euclidean "no boundary" $S^3$ sphere which is analytically continue into Lorentzian spatially closed de-Sitter universe. Omit the $l$ and $m$ index, we get Euclidean equation of motion
%\begin{equation} \label{peome}
%	\hat{f}''_n+3H\hat{f}'_n-((\frac{\partial \tilde{V}}{\partial\phi})^2+\frac{n^2-1}{a^2})\hat{f}_n=0,
%\end{equation}
%and Lorentzain equation of motion
%\begin{equation} \label{peoml}
%	\ddot{f}_n+3H\dot{f}_n+((\frac{\partial \tilde{V}}{\partial\phi})^2+\frac{n^2-1}{a^2})f_n=0.
%\end{equation}

In order to impose the regularity condition for $f_{n}$ at the south pole $\tau = 0$, we need to set $f_{n}(\tau = 0)$ for $n \geq 2$ or $df_{n}/d\tau (\tau = 0) = 0$ for $n = 1$ \cite{Halliwell1990}. For $n \geq 2$, we introduce the initial condition
\begin{eqnarray}
	f_n(\tau_i) &=& \frac{1}{2}\epsilon\tau_i^2, \\
	\dot{f}_n(\tau_i) &=& \epsilon\tau_i,
\end{eqnarray}
where $\tau_i\ll 1$ is the initial Euclidean time and $\epsilon$ is an arbitrary small parameter. Since the expectation value $\langle f_n^2\rangle$ depends only on the ratio $\tilde{f}_n/\dot{\tilde{f}}_n$, the power spectrum is independent of the choice of $\epsilon$. The freedom to choose the initial condition is constraint by the regularity condition. Note that there exists two linearly independent solutions for $f_{n}$, one is regular and the other is singular at $a = 0$. However, since the singular solution rapidly damped out for large $\tau$, if we give such a set of initial conditions, then we obtain the correct results for large $\tau$.

In our numerical calculations, we set $\tau_i=10^{-4}$ and $\epsilon=10^{-4}$. At the turning time from Euclidean to Lorentzian signatures, we impose the following matching condition \cite{Hwang2012}
\begin{eqnarray}
	\textrm{Re} f_n|_{t = 0} &=& \textrm{Re} f_n|_{\tau = \frac{\pi}{2H_0}}, \\
	\textrm{Im} f_n|_{t = 0} &=& \textrm{Im} f_n|_{\tau = \frac{\pi}{2H_0}}, \\
	\textrm{Re} \dot{f}_{n}|_{t = 0} &=& -\textrm{Im} \dot{f}_{n}|_{\tau = \frac{\pi}{2H_0}}, \\
	\textrm{Im} \dot{f}_{n}|_{t = 0} &=& \textrm{Re} \dot{f}_{n}|_{\tau = \frac{\pi}{2H_0}}.
\end{eqnarray}
Although $f_{n}$ is complex valued in general, since all equations are linear, one can always make $f_{n}$ to be real at the horizon crossing time by introducing a phase factor (i.e., introducing a proper $\epsilon$.) Therefore, without loss of generality, we can calculate the power spectrum by using this $f_{n}$.

%For each mode $n$, we compute $P(n)$ at the horizon exit $k=aH$. In the following paragraph we will show the numerical results of $P(n)$ with two types of potential, Starobinsky and modified Starobinsky. Then we will develop the relation with $P(n)$ and CMB spectrum and show that with certain choice of parameter, the result better fit with CMB data.

%%%%%%%%%%%%%%%%%%%%%%%%%%%%%%%%%%%%%%%%%%%%%%%%%%%%%%%
\subsection{Starobinsky model}

Now we adopt the techniques we mentioned earlier to the Starboinsky-type inflation model. The action of the Starobinsky model can be written as \cite{Starobinsky1980}
\begin{equation}
	S=\frac{1}{16\pi}\int d^4x \sqrt{-g} \left( R+\frac{1}{6M^2}R^2 \right),
\end{equation}
where $M$ is a constant. This is equivalent to the following model in the Einstein frame by applying conformal transformation:
\begin{equation}
	S=\int d^4x \sqrt{-g} \left[ \frac{R}{16 \pi} - \frac{1}{2} \partial^\mu{\phi}\partial_\mu{\phi}-V_{0}\left(1-e^{-\sqrt{\frac{16 \pi}{3}}\phi}\right)^2 \right],
\end{equation}
where $V_{0} = 3M^2/32\pi$. By choosing $\sigma^{2} = 1/V_{0}$, one can see that the essential dynamics of the solution is independent of $\sigma$, while the only probability has the overall dependence of $\sigma$.

\begin{figure}
\begin{center}
\includegraphics[scale=0.5]{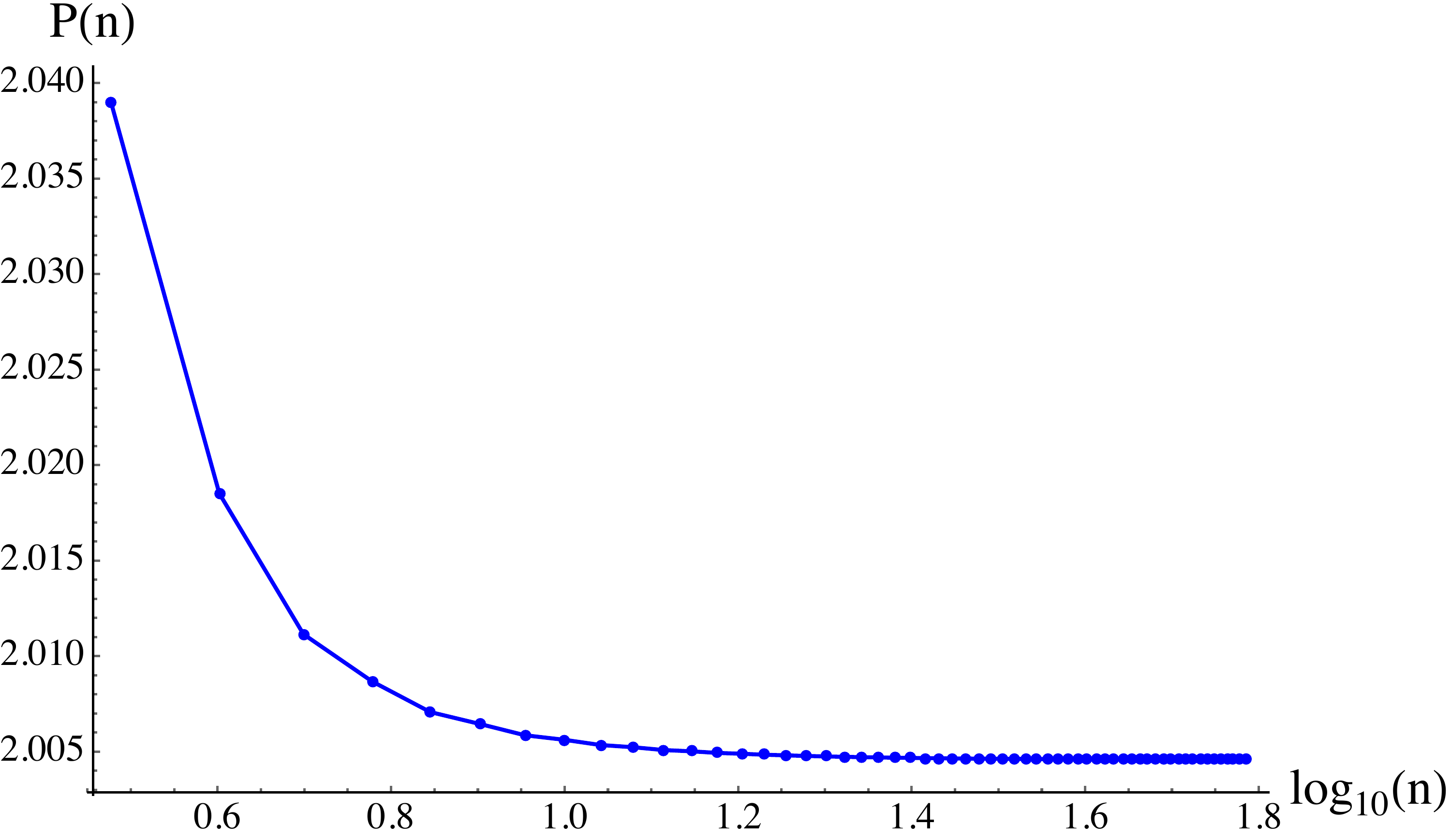}
\caption{\label{fig:V_0=1}The power spectrum of the original Starobinsky model as a function of the mode number $n$. The spectrum is enhancement at the small $n$ limit, while it satisfies the scale invariance at the large $n$ limit.}
\end{center}
\end{figure}

The result of the power spectrum $P(n)$ is shown in Fig.~\ref{fig:V_0=1}. It is easily to notice that the power spectrum is enhanced for the large scales \cite{Starobinsky1996}, while it shows the scale-invariance for the small scales as we expected \cite{Halliwell1985}. This justifies that the Euclidean vacuum condition reproduces the standard BD vacuum.

%this theory is simply a minimal coupled scalar field with plateau like potential $V(\phi)=\frac{3}{4}\frac{M^2}{\kappa^2}(1-e^{-\sqrt{\frac{2\kappa^2}{3}}\phi})^2=V_0(1-e^{-\sqrt{\frac{2\kappa^2}{3}}\phi})^2$. It is not difficult to show that, the eqaution \eqref{peoml} only depends on $\frac{V}{V_0}$ while the metric rescales as $\sigma^2=\frac{1}{V_0}$. The physical reason is justified because the potential V only tells you the energy scale of the inflaton, in other words, while the full power spectrum should depend on the amplitude of $V_0$, the precise value of $V_0$ is irrelevant for our instanton solution of perturbed scalar field. Without loss of generality, we choose $V_0=1$ and there is no free parameter in the calculation. 

%%%%%%%%%%%%%%%%%%%%%%%%%%%%%%%%%%%%%%%%%%%%%%%%%%%%%%%
\subsection{Modified Starobinsky model}

\begin{figure}
\begin{center}
\includegraphics[scale=0.75]{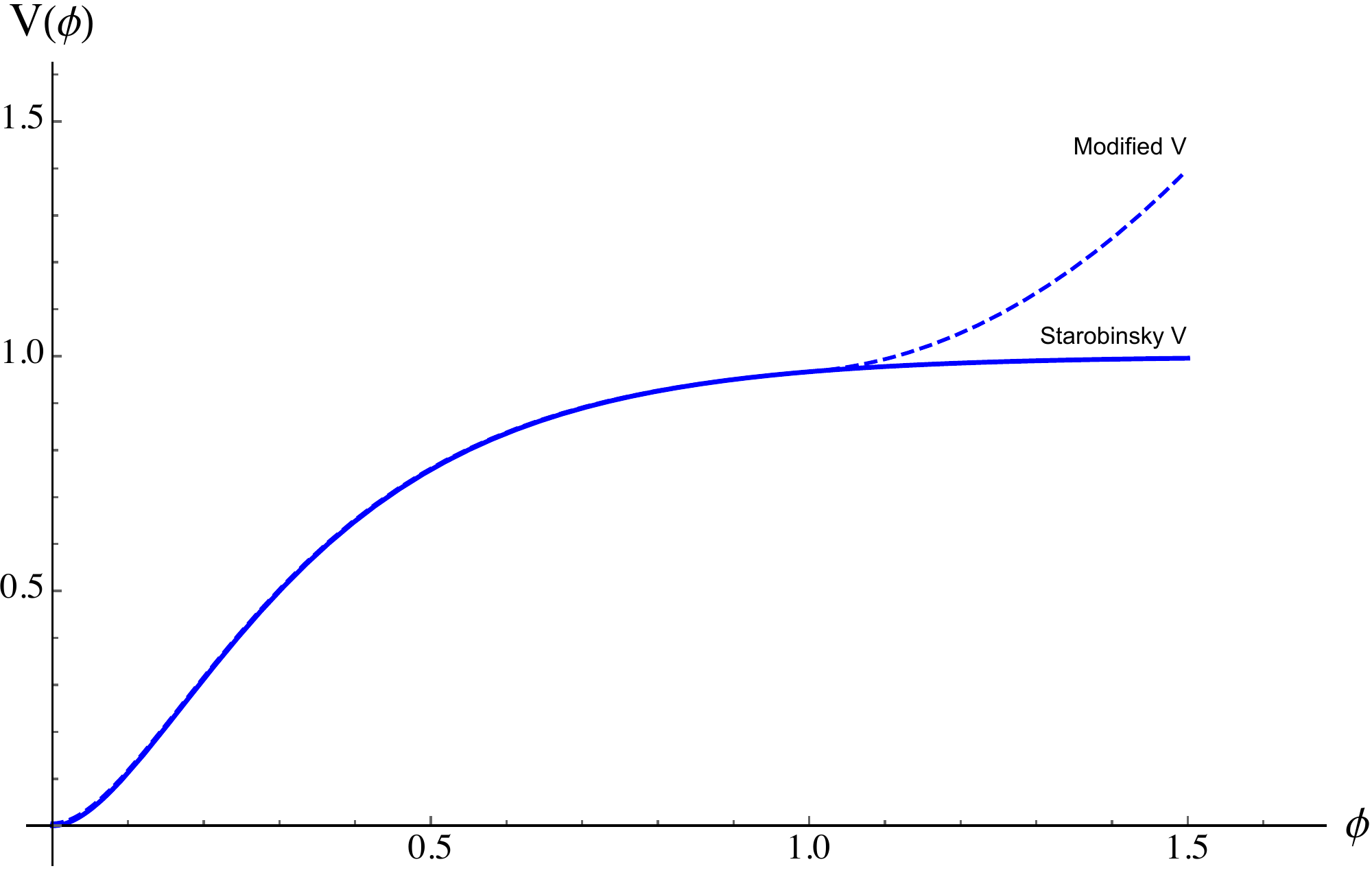}
\caption{\label{fig:Vplot}Comparison of the inflaton potential between the original Starobinsky model and its modified version.}
\end{center}
\end{figure}

In the above paragraph, we shown that the Starobinsky model does not allow power suppression for the long-wavelength modes. However, as we observed in \cite{Chen:2017aes}, if the effective mass term is applied for the earlier stage of inflation, there is a hope to see the power suppression. Hence, we modify a little bit of the Starobinsky model as follows:
\begin{equation} \label{eq:aperp} 
	V=V_0 \left[ \left(1-e^{-\sqrt{\frac{2\kappa^2}{3}}\phi}\right)^2+\frac{1}{2}\mu^2(\phi-\phi_0)^2 \left( \frac{1}{2}+\frac{1}{\pi}\arctan{\frac{\phi-\phi_0}{\Delta}} \right) \right],
\end{equation}
where $\mu$ characterize the mass scale (i.e., $\mu^{2} = m^{2} / V_{0}$), $\Delta$ and $\phi_{0}$ are model parameters,  see Fig.~\ref{fig:Vplot}. At the first glimpse, this model is just designed for the power suppression, but theoretically there are several reasons to consider this kind of model (see Appendix A). The meaning of each parameters are explained in the follows, $\Delta$ is a parameter for a smooth connection between the Starobinsky model $\phi < \phi_{0}$ and the pre-inflationary stage $\phi > \phi_{0}$, and hence we choose a small value, e.g., $\Delta=10^{-7}$ in order to assure that the region $\phi < \phi_0$ is unchanged for numerical calculations. $\phi_0$ is a value for the end of the pre-stage inflation, and we choose $\phi_0$ such that the inflation after the pre-inflation gives enough $e$-foldings, e.g., approximately $60$ $e$-foldings; this implies that $\phi_{0} \simeq 1.1$ becomes a good choice. Then, the effects of the pre-inflationary stage will only contribute to the long-wavelength modes, where the scale invariance is preserved at small scales as we expected.

\begin{figure}
\begin{center}
\includegraphics[scale=0.5]{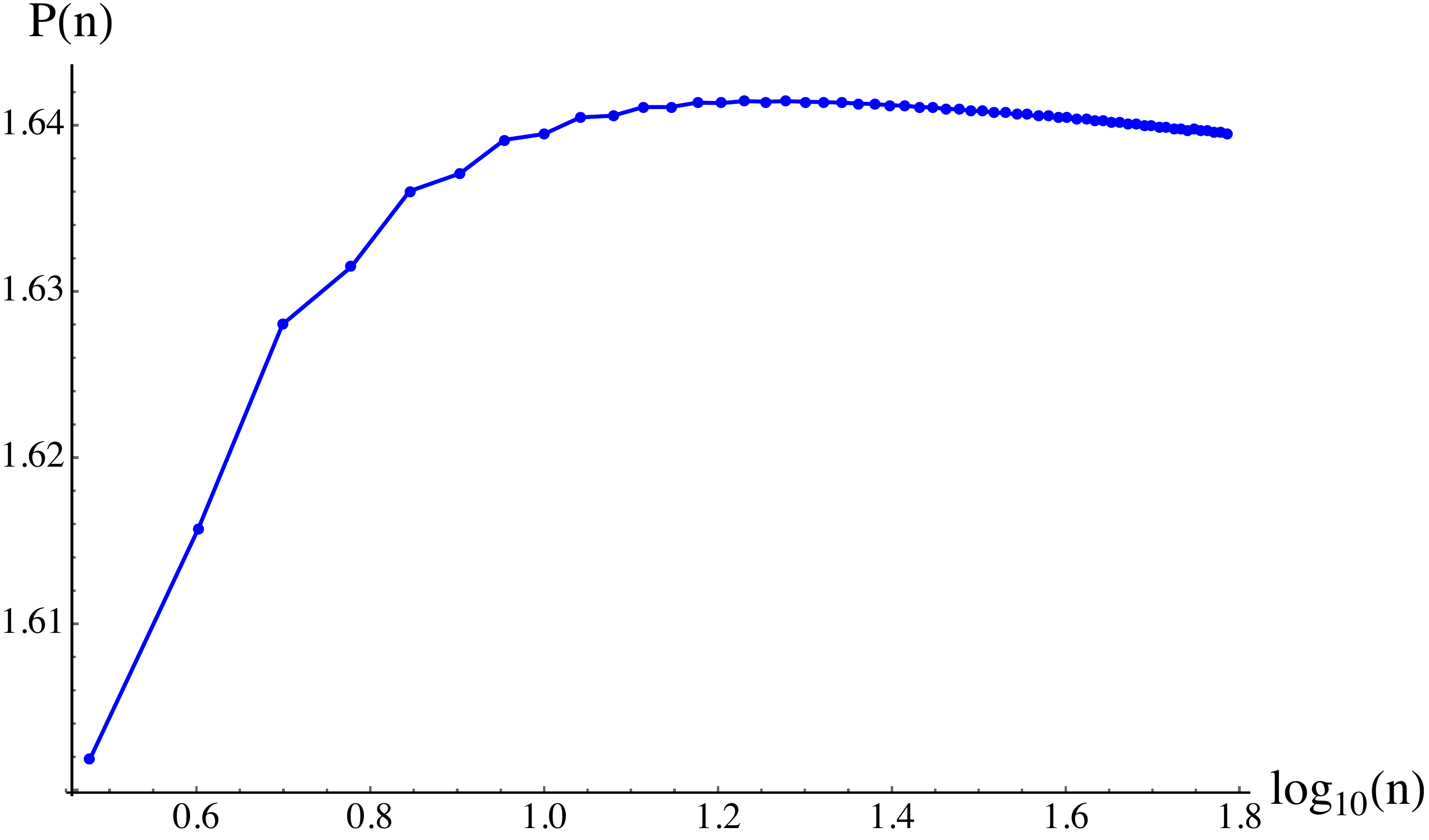}
\caption{\label{fig:mu=18x0=11}The power spectrum of Starobinsky-type potential in terms of mode $n$ with $\mu=1.8$, $\phi_0=1.1$. The spectrum is suppressed at small $n$, while remains scale invariant at large $n$.}
\end{center}
\end{figure}

Our numerical simulation shows that a moderate amount of the mass parameter $\mu$ is sufficient to account for the power suppression. As an example, with $\mu=1.8$, $\phi_0=1.1$, one can see the significance suppression at the large scale, see Fig~\ref{fig:mu=18x0=11}. Note that in order to satisfy the classicality of the background instanton solution, the constraint $\mu \leq \sqrt{6\pi} \simeq 4.34$ should be satisfied \cite{Hwang2013a}.

\begin{figure}
\begin{center}
\includegraphics[scale=0.27]{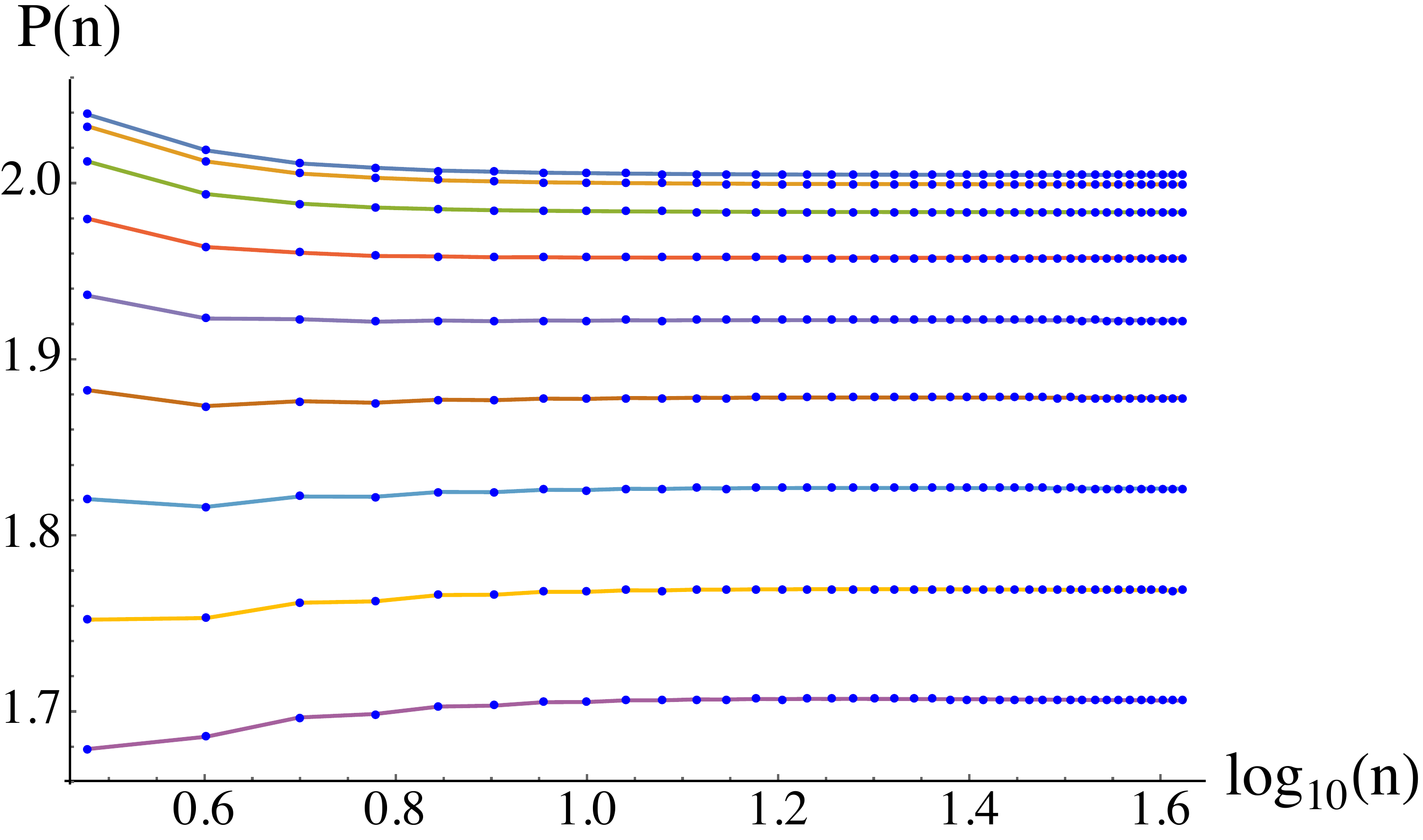}
\includegraphics[scale=0.27]{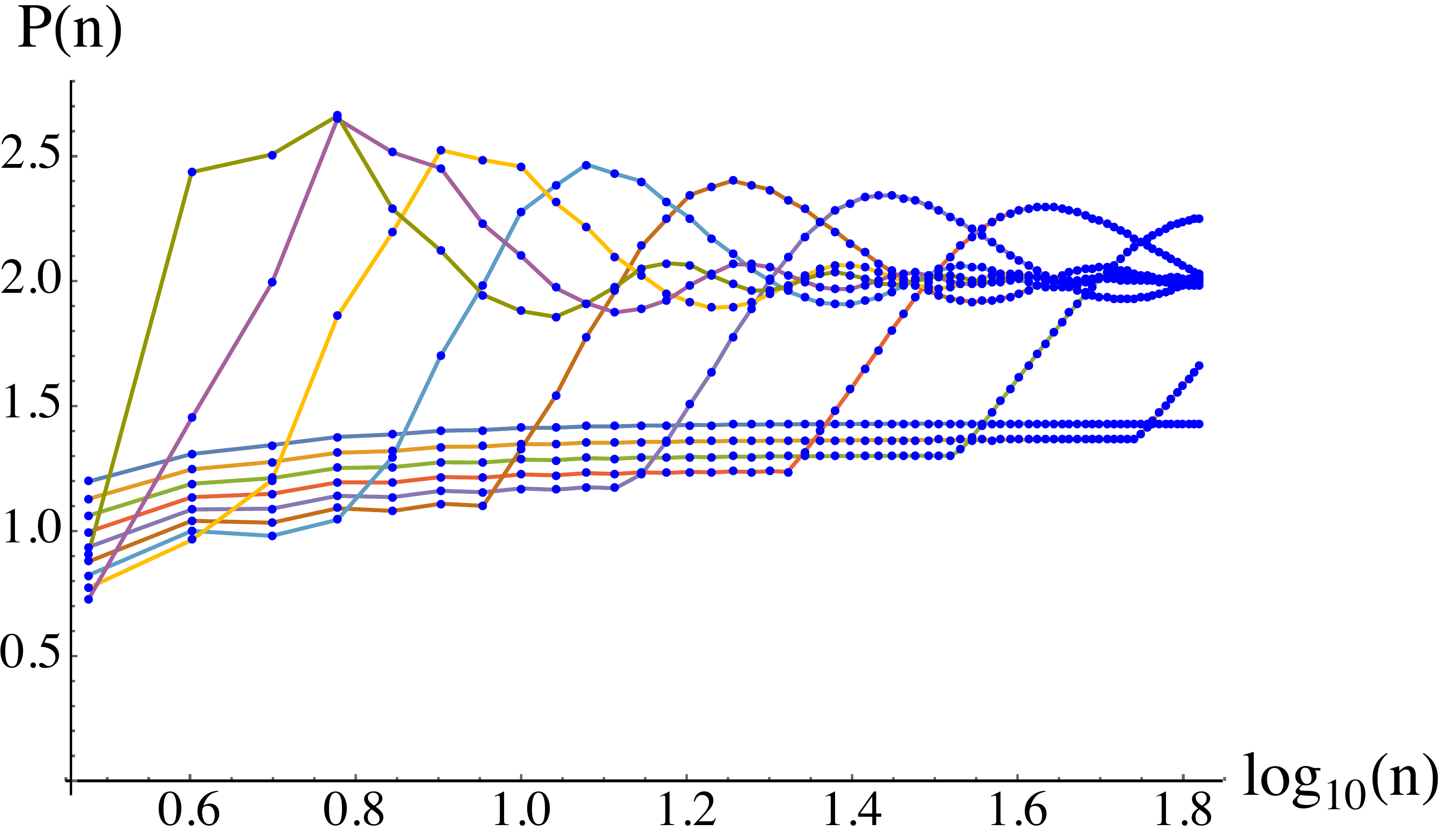}
\caption{\label{fig:mu=0-2x0=11}Left shows the power spectrum by varying $\mu$ with the same $\phi_0 = 1.1$. From top to bottom, $\mu$ varies from $0.2$ to $2$ with the step size $0.2$. The spectrum turns from enhancement to suppression for the large scales. Right shows the power spectrum for $\phi_0 = 1.1$ and different $\mu$ from $2.4$ to $4.2$ with the step size $0.2$. There appears a steep suppression on large scales with oscillatory behaviors for the small length scales. (Note that $\mu < \sqrt{6\pi}$ is required for the existence of classicalized instantons.)}
\end{center}
\end{figure}

\begin{figure}
\begin{center}
\includegraphics[scale=0.7]{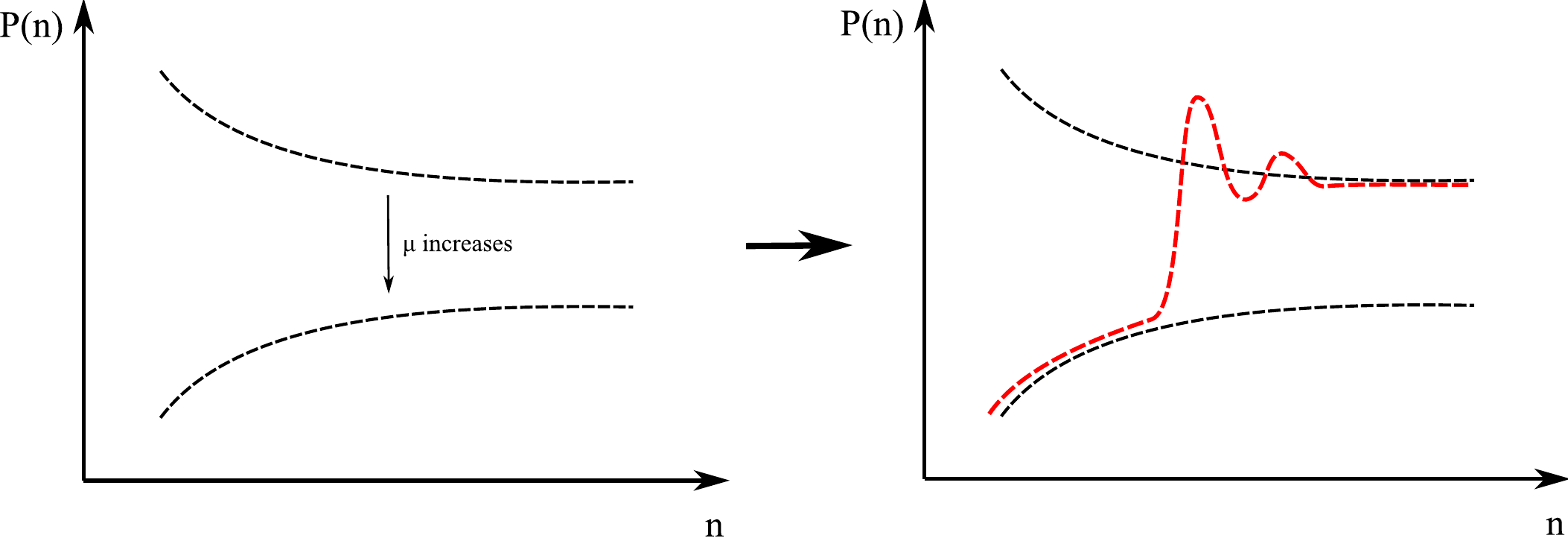}
\caption{\label{fig:match}We illustrate the mechanism to induce a steep suppression (red curve) by introducing a pre-stage inflaton potential. Here the spectrum is connected by the matching on the transition point $\phi\approx\phi_0$.}% and the suppression is introduced by the lack of overall amplitude of $H$ in the early $\phi>\phi_0$ stage.}
\end{center}
\end{figure}

To see the generic behavior of the power spectrum, we fix $\phi_0 = 1.1$ and vary $\mu$ from $0$ to $2$, see Fig.~\ref{fig:mu=0-2x0=11}. If we increase $\mu$, the tendency of the large scale mode changes form enhancement to suppression, while this is consistent with our previous result in the $\phi^2$ model \cite{Chen:2017aes}. However, this is not sufficient to explain the power suppression of the large length scales since the amount of suppression is too weak. Surprisingly, if we set $\mu > 3$, it occurs that there is a steep suppression at large scales (right of Fig.~\ref{fig:mu=0-2x0=11}), which accounts for the large scale anomaly in CMB spectrum. If we increase the slope of the inflaton potential, i.e. increase $\mu$, the spectrum decrease its overall amplitude. Then the physics is clear that when introducing the pre-stage inflation, the sudden change in the amplitude will result in the power suppression at the large scale. The similar result was obtained by introducing a kinetic energy dominated stage before the inflation \cite{Contaldi:2003zv}. We summarized the tendency in Fig.~\ref{fig:match} as an illustration.

%A natural question is that \textit{"Why there is a suppression at low modes if we introduce the pre-stage inflation?"} Or conversely, \textit{"Why not in the original Starobinsky model?"} Here, we find out the similar result which was obtained by introducing a kinetic energy dominated stage before the inflation \cite{Contaldi:2003zv}. The authors observed that by increasing the slope of the potential, the spectrum is suppressed. If the scenario is correct, then we may interpret the suppression as the consequences of the matching between two different slope of the potential. As a result, the oscillation happens naturally as the connection of the two stages.
%r

%\begin{figure}
%\begin{center}
%\includegraphics[scale=0.4]{loga}
%\includegraphics[scale=0.4]{dotphi}
%\includegraphics[scale=0.4]{H2}
%\includegraphics[scale=0.4]{gaugefactor}
%\caption{\label{fig:loga}The background evolution in terms of the Hubble time $t$. Upper left shows that there are around $60$ number of $e$-foldings. Upper right and lower left show the evolution of $\dot{\phi}$ and $H^2$, while the Hubble parameter rapidly increases at the early stage of inflation and then turns into the quasi de-Sitter expansion phase. Lower left shows the evolution of $(H/\dot{\phi})^2$.}
%\end{center}
%\end{figure}

%%%%%%%%%%%%%%%%%%%%%%%%%%%%%%%%%%%%%%%%%%%%%%%%%%%%%%%
\subsection{Connection with the CMB spectrum}

\begin{figure}
\begin{center}
\includegraphics[scale=0.27]{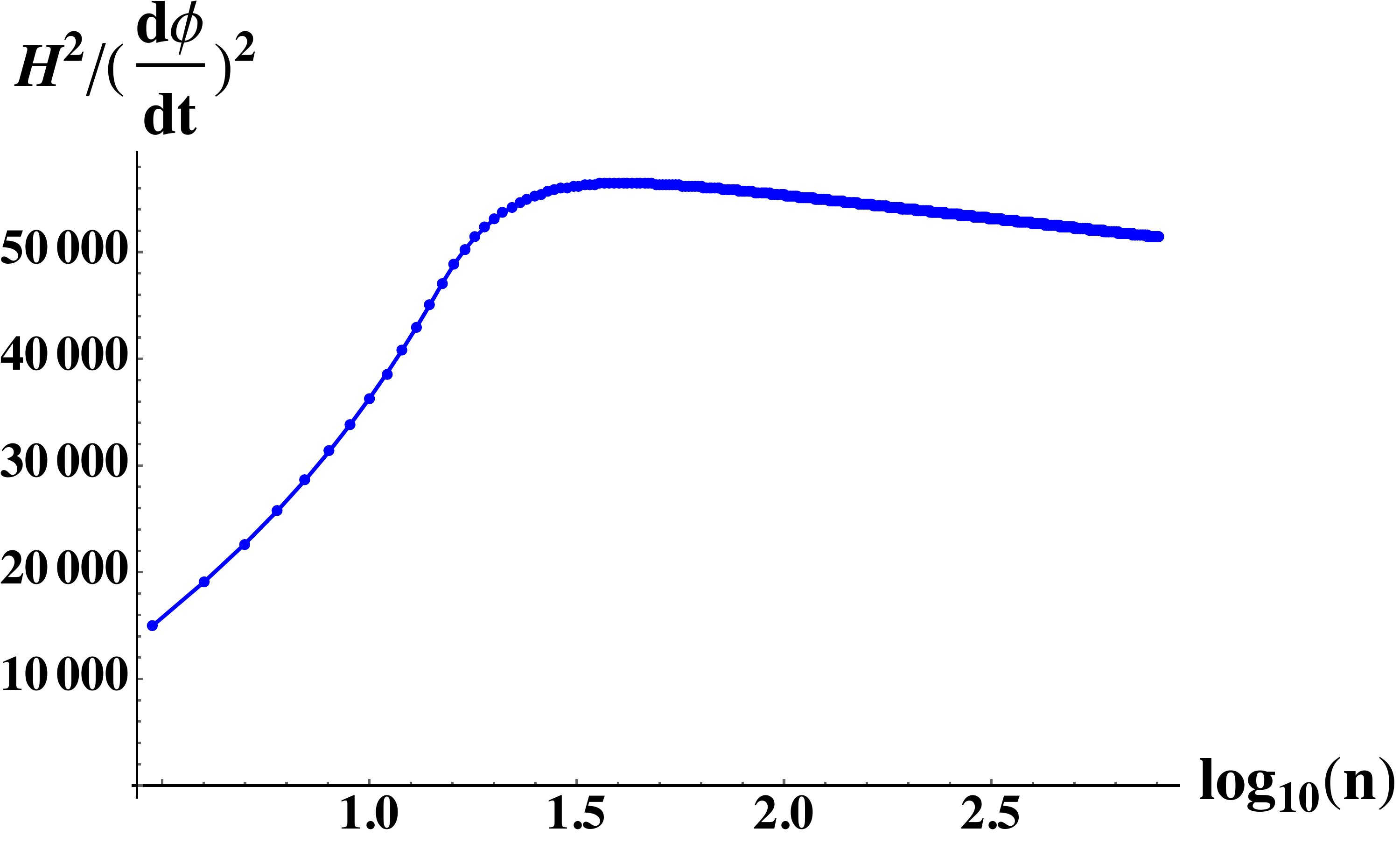}
\includegraphics[scale=0.27]{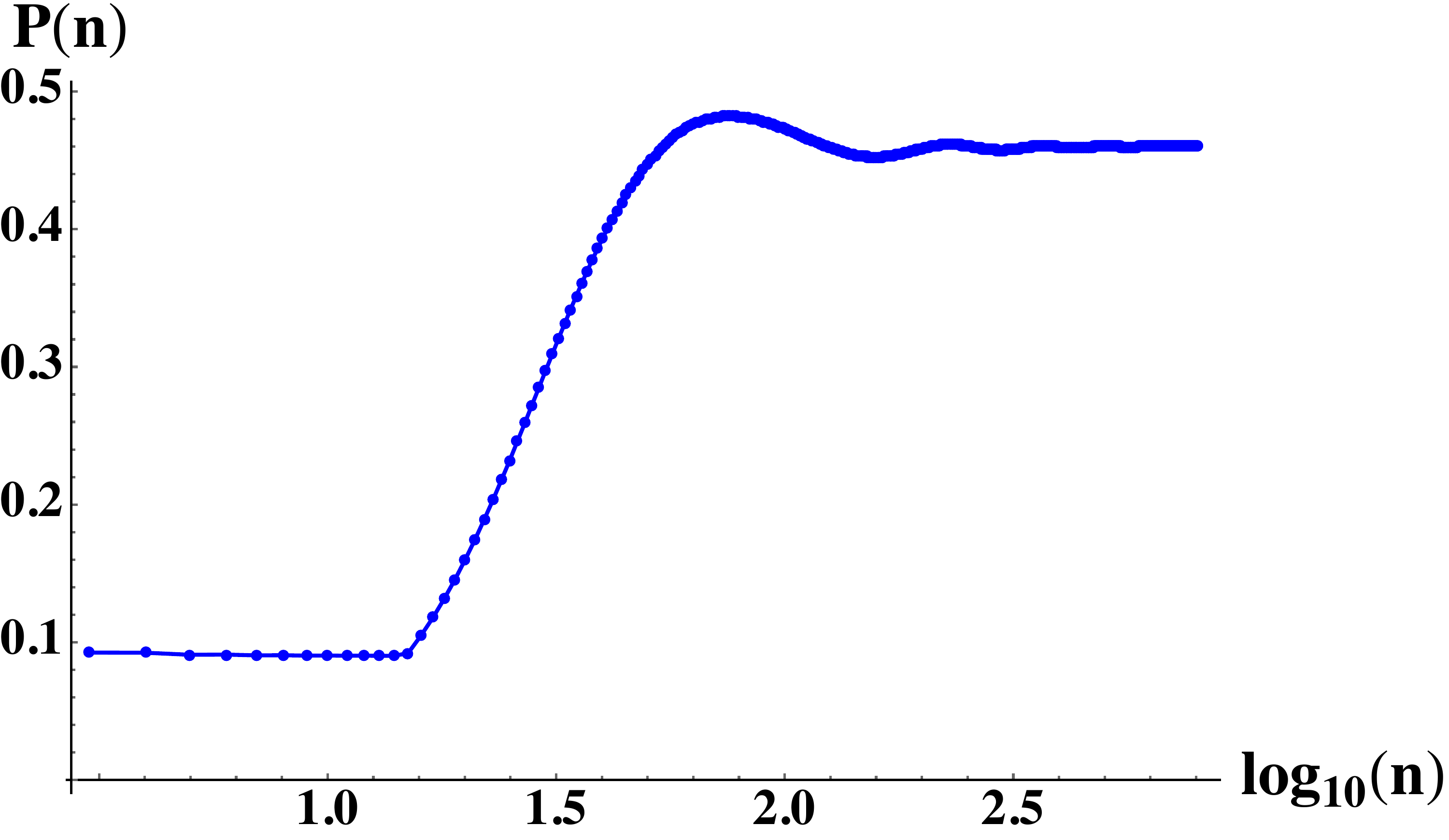}
\caption{\label{fig:gaugen}Left shows that the background part indeed has the effect of the power suppression at large scales with parameter $\mu=2.7$, $\phi_0=1.07$, and $\phi(t=0)=1.123$, while right shows the steep suppression due to the perturbation part.}
%\end{center}
%\end{figure}
%\begin{figure}
%\begin{center}
\includegraphics[scale=0.5]{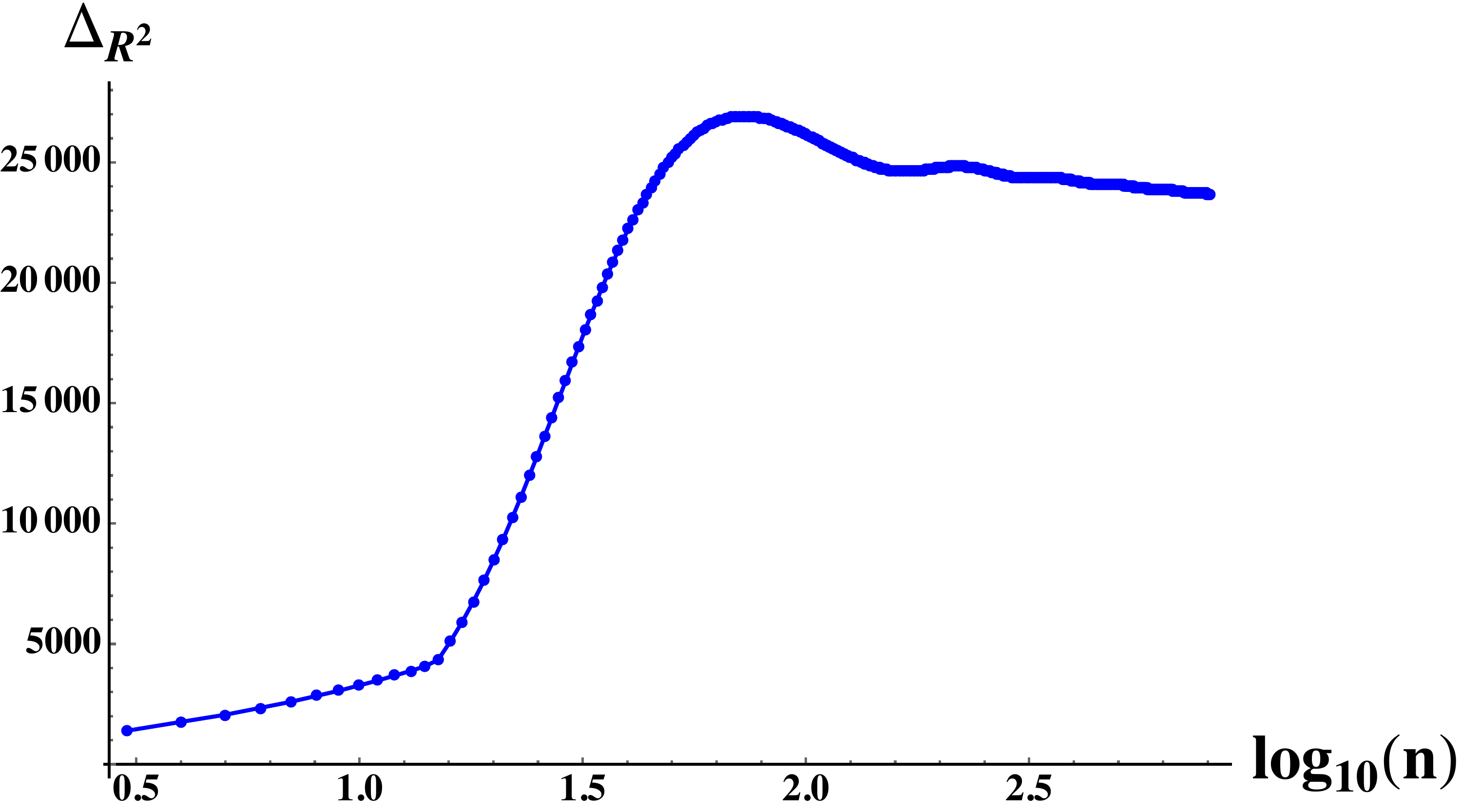}
\caption{\label{fig:Deltan}The gauge invariant primordial power spectrum combining two plots in Fig.~\ref{fig:gaugen}.}
\end{center}
\end{figure}

With the power spectrum $P(n)$, we can establish the correspondence between the $P(n)$ and the CMB spectrum. Note that P(n) is not gauge invariant, the correct interpretation of the primordial power spectrum should be written in the gauge invariant form \cite{Gratton:2001gw}. One of the gauge invariant object that we are usually chosen is 
\begin{equation}
	 \mathcal{R}=\Psi+\frac{H}{\dot{\phi}}\delta\phi.
\end{equation}
Where $\mathcal{R}$ is comoving curvature perturbation and $\Psi$ is gravitational potential. The power spectrum can be calculated by Fourier transform the real space correlation function of $\mathcal{R}$. Which is by definition
\begin{equation} 
	\langle\mathcal{R}\mathcal{R}\rangle=\sum_n\frac{n}{n^2-1}\Delta^2_\mathcal{R}(n).
\end{equation}
Where $\Delta^2_\mathcal{R}(n)$ is the gauge invariant power spectrum. By making a gauge choice $\Psi=0$, the primordial power spectrum is simply
\begin{equation} 
	\Delta^2_\mathcal{R}(n)= \left( \frac{H}{\dot\phi} \right)^2 \frac{3n(n^2-1)}{8\pi\sigma^2 a^3}\frac{\bar{f_n}}{\dot{\bar{f_n}}}= \left( \frac{H}{\dot\phi} \right)^2 P(n),
\end{equation}
with the variables evaluated at the horizon crossing time. The spectrum can be divided by two parts, where $(H/\dot{\phi})^2$ describes the kinetic rolling feature of the background while $P(n)$ describes the perturbed scalar field. %For example, in Fig.~\ref{fig:bgevo}, we can see that in the earlier stage of the inflation, the evolution of $(\frac{H}{\dot{\phi}})^2$ does not admit the slow-roll feature, while after the turning point $t \approx 2$, the evolution turns back to the slow-roll limit. It is not difficult to see that these strange features are due to the choice of the matching condition $\dot{a}(t=0)=0$, where we will mention this point in the discussion section. 
In Fig.~\ref{fig:gaugen}, it is shown that, for $\mu=2.7$, $\phi_0=1.07$, and $\phi(t=0)=1.123$, the behavior of the background $(H/\dot{\phi})^2$ and perturbation $P(n)$ both have a steep suppression. Combining these results, the gauge invariant spectrum is shown in Fig.~\ref{fig:Deltan}. Note that by choosing the Hartle-Hawking's no boundary state, it requires $H=0$ at the Wick rotation. Thus no matter what spectrum $P(n)$ is, the background term $(H/\dot{\phi})^2$ always have strong suppression at large scales.

\begin{figure}
\raggedleft
\includegraphics[scale=0.58]{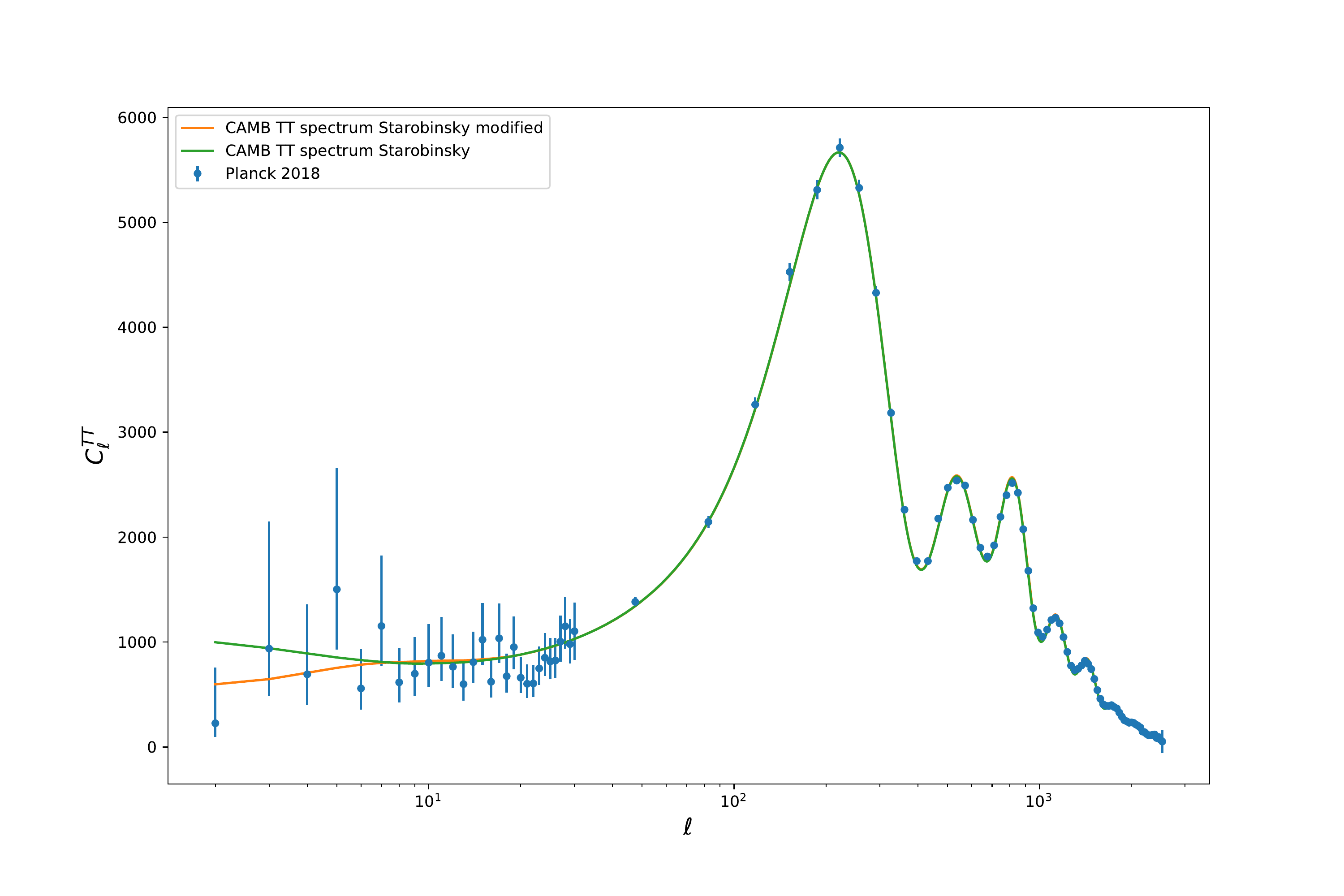}
\caption{\label{fig:CMB}This figure shows the result of the CAMB calculation. The orange curve is the result using modified Starobinsky model with the parameters $\mu=2.7$, $\phi_0=1.07$, and $\phi(t=0)=1.123$, which shows the suppression for the large scale modes, while the green curve is the result predicted by the original Starobinsky model.}
\end{figure}

For each mode $n$, there is a relation with the comoving wavenumber $k$:
\begin{equation} \label{rela}
	k^2=\frac{n^2-1}{R^2_c},
\end{equation}
where $R_c$ is the comoving radius of the $S^3$ sphere and relates to the curvature density of the universe via
\begin{equation}
	R_c=\frac{|\Omega_k|^{-\frac{1}{2}}}{aH}.
\end{equation}
That is to say, for an extremely flat universe $|\Omega_k|\rightarrow 0$, the comoving wavenumber of large scale modes will be too small to see any effects within the current observation. According to Planck 2018 data, the upper bound of the curvature density is $\Omega_k = 0.0007 \pm 0.0019$ (with Planck TT, TE, EE + lowE + lensing + BAO data within 68 $\%$ confidence level.) Within the observational bound, we assume that $\Omega_k = -0.0012$ so as not to make the universe too large. Together with the current Hubble parameter $H_0=67.4$ km $\text{s}^{-1}$Mp$c^{-1}$, we can calculate $R_c=128.2$ Gpc. On the contrary, the largest observable scale is given by the last scattering surface (here we assume the recombination is instantaneous), $r_{ls}=14.0$ Gpc, where this gives the comoving wavenumber $k_{ls}={2\pi}/2{r_{ls}}$. We can estimate the observable modes by using $R_c$, $k_{ls}$, and Eq.~\eqref{rela}; then we get $n \simeq 28$. This means that the primordial spectrum with modes $n > 28$ are observable in the CMB power spectrum.

With these information, we are able to construct the CMB power spectrum. It is not easy to directly translate the primordial power spectrum into the CMB anisotropy spectrum since there are several effects that need to be considered, e.g., the Sachs-Wolfe effect and the Sunyaev-Zel'dovich effect. In order to do this, we used the CAMB code \cite{Bonga:2016iuf} to calculate the CMB spectrum. In Fig.~\ref{fig:CMB}, we show that the spectrum is better fitted at the low $\ell$ modes in the $C^{TT}_\ell$ spectrum. While the quadrupole anomaly is alleviated, the $\ell = 20 \sim$ 30 anomaly is not affected by this effect. More detailed relations between the primordial spectrum and the observational CMB spectrum are shown in Appendix B.

%%%%%%%%%%%%%%%%%%%%%%%%%%%%%%%%%%%%%%%%%%%%%%%%%%%%%%%
\section{\label{sec:conclu}Conclusion}

In this paper we addressed the low multipole anomaly issue in the CMB power spectrum. We invoked Hartle-Hawking's no boundary proposal to construct the power spectrum. We found that if the underlying spacetime has a nonzero curvature, then the perturbations at large scales may deviate from that under scale-invariance. In order to calculate the power spectrum from the Euclidean path integral approach, we used the steepest-descent approximation and the Euclidean vacuum condition. This allowed us to calculate the expectation value of the power spectrum on the fix background geometry. We further analyzed the power spectrum with the Starobinsky model and found that the the matter perturbations are enhanced rather than suppressed on large scales. However, when we introduced a pre-stage inflation, we showed that it is possible to simultaneously suppress the power spectrum at large scales and preserve the scale-invariance at small scales.

There are two reasons for such power suppression at large scales. First, given the power spectrum $\Delta \sim H/\dot{\phi} \times \delta \phi$, we found that $\delta \phi$ is suppressed at large length scales because of our introduction of the pre-inflation stage.  (that is consistent with \cite{Contaldi:2003zv,Yamauchi:2011qq} though the detailed mechanism is different). Second, this suppression is amplified due to the behavior of $H$, which is approximately proportional to the Hawking temperature. We note that the effective Hawking temperature is zero soon after the Wick-rotation. Hence perturbations during the pre-inflation stage must be significantly suppressed. 

%The introduction of pre-stage inflation is similar to the approach of \cite{Contaldi:2003zv,Yamauchi:2011qq} that introduced a kinetic energy dominated stage before the premordial inflation. We have shown that the spectrum is recovered in the large $n$ and small $\mu$ limit (the original Starobinsky model). However, we point out that there is a difference between the canonical approach of calculating the power spectrum and our approach. In the standard approach, the vacuum is not well defined. The usual assumption is that all of the modes are generated in Minkowski space at conformal time $\eta=-\infty$, this is what we called Bunch Davies vacuum. However, this is only valid for the modes which is well inside the Hubble  horizon at the onset of inflation. If the duration of inflation is just about 60 e-folds, the large scale mode will be observable to us. Therefore the behavior of these modes will strongly depend on the initial condition before inflation, and the ambiguity of the vacuum selection is recovered. In this sense, our approach is better since we have a well-defined vacuum state, which is a no boundary wave function. The observable is directly connect with the theory itself.

There are several important issues that require further investigations. First, in this work we only restricted our interests to the $C^{TT}_{\ell}$ spectrum, but but it should be extended to other components. Second, our framework is based on Einstein's general relativity augmented by Hartle-Hawking insrtantons, but there are other options to follow. For example, one may consider including non-perturbative quantum gravitational \cite{Brahma:2018elv} or considering non-compact instantons \cite{Chen:2016ask}. These additional instantons will affect the power spectrum and may provide additional experimental tests to quantum gravity effects.

%%%%%%%%%%%%%%%%%%%%%%%%%%%%%%%%%%%%%%%%%%%%%%%%%%%%%%%
\section*{Acknowledgment}
The authors would like to thank Yu-Hsiang Lin for fruitful conversations. We also thank to Frederico Arroja, Rio Saitou, Jiro Matsumoto, Fabien Nugier, and other fellows of LeCosPA. PC and HY are supported by Taiwan National Science Council under Project No. NSC 97-2112-M-002-026-MY3, Leung Center for Cosmology and Particle Astrophysics (LeCosPA) of National Taiwan University, and Taiwan National Center for Theoretical Sciences (NCTS). PC is in addition supported by US Department of Energy under Contract No. DE-AC03-76SF00515. DY is supported by the Korean Ministry of Education, Science and Technology, Gyeongsangbuk-do and Pohang City for Independent Junior Research Groups at the Asia Pacific Center for Theoretical Physics and the National Research Foundation of Korea (Grant No.: 2018R1D1A1B07049126).

%%%%%%%%%%%%%%%%%%%%%%%%%%%%%%%%%%%%%%%%%%%%%%%%%%%%%%%
\section*{Appendix A: Preference of large $e$-foldings in the no-boundary proposal}

One typical problem of the Hartle-Hawking wave function is that it does not prefer a large number of $e$-foldings. Around the local minimum, let us approximate he potential as
\begin{eqnarray}
V(\phi) \simeq V_{0} \left( 1 + \frac{1}{2} \mu^{2} \phi^{2} \right).
\end{eqnarray}
Let us assume that $\phi(t = 0) \equiv \varphi$ at the turning point. Then according to \cite{Hartle2008a}, the probability is
\begin{eqnarray}
P\left[ \varphi \right] \propto \exp{\frac{3}{8V(\varphi)}}.
\end{eqnarray}
However, in order to satisfy the classicality of the inflaton field, if $\mu \geq \sqrt{6\pi}$, there exists classicalized instantons only if $\varphi \gtrsim \varphi_{0}$. Here, $\varphi_{0}$ depends on $\mu$ and for the $\mu \gg 1$ limit, $\varphi_{0} \sim 0.62$ \cite{Hwang:2012bd}. (If $\mu < \sqrt{6\pi}$, then for any $\varphi$, there exists classicalized instantons.) Therefore, based on \cite{Hartle2008a}, the most preferred classicalized instanton should be $\varphi \sim \varphi_{0}$ which results the $e$-foldings $N \simeq 2\pi \varphi_{0}^{2} \simeq 2.4$, while our universe requires around $50$ $e$-foldings and this is exponentially disfavored.

In order to overcome this problem, there have been several trials \cite{Hwang2014}.
\begin{itemize}
\item[--] 1. \textit{Volume weighting}: In \cite{Hartle2008a}, the authors introduced so-called the top-down approach. This means that in order to give the correct measure of the wave function, one needs to count not only the bottom-up factor (wave function) but also the volume factor. So, the probability is written in terms of final $e$-folding $N$ by \cite{Hwang:2012bd}
\begin{eqnarray}
P\left[ N \right] \propto \exp{\left(\frac{3\pi}{2 m^{2} N} + 3 N\right)}.
\end{eqnarray}
Therefore, the volume factor can be used to favor large $e$-foldings. However, there is no fundamental justification for this approach.
\item[--] 2. \textit{Possible ideas of \cite{Hwang2014}}: In \cite{Hwang2014}, the authors discussed that there are three possible ways to make the model to prefer large $e$-foldings (assuming Einstein gravity and effectively a single scalar field).
\begin{itemize}
\item[--] 2-1. If the pre-inflation (and the beginning of the universe) is started at the Planck scale, then the exponential hierachy would be milder than the current inflationary scenario.
\item[--] 2-2. If the potential is very highly tuned such that the cutoff scale corresponds enough $e$-foldings, then it can explain large $e$-foldings.
\item[--] 2-3. If there is a new factor that increases the phase space for the large $e$-folding regime, e.g., the large field space or large number of assisted fields \cite{Hwang:2012bd}, then it can explain large $e$-foldings.
\end{itemize}
All of these scenarios are theoretically possible, but require many ad hoc fine-tunings.
\item[--] 3. \textit{Modified gravity}: For example, the dRGT massive gravity \cite{Sasaki:2013aka} or bi-gravity \cite{Zhang:2014wia} can explain the large $e$-foldings. However, these models go beyond the Einstein gravity.
\end{itemize}

On the other hand, there may be a fourth way \cite{Hwang2015}: if one goes beyond the single field model with Einstein gravity, then one can get a simpler explanation. Let us assume that there are two fields:
\begin{eqnarray}
V(\phi_{1},\phi_{2}) = V_{1}(\phi_{1}) + \frac{1}{2} m_{2}^{2} \phi_{2}^{2},
\end{eqnarray}
where $V_{1}(\phi_{1})$ is the usual inflaton field potential that satisfies the slow-roll condition and $\phi_{2}$ is an additional field with relatively large $m_{2}$. In the beginning of the universe, two fields should be classicalized at the same time. In order to do this, one needs to satisfy
\begin{eqnarray}
\frac{m_{2}^{2}}{V_{1}(\phi_{1}(0))} < 6 \pi,
\end{eqnarray}
which is nothing but a generalization of the classicality condition for a single scalar field \cite{Hartle2008a}. In other words, $V_{1}(\phi_{1}(0)) > m_{2}^{2} / 6\pi$. By suitably choosing $m_{2}$, one can increase the cutoff for the inflaton field direction. This requires large $V_{1}(\phi_{1}(0))$, or equivalently large $e$-foldings. In the end, from numerical computations, one can estimate the most preferred initial condition and indeed it prefers enough $e$-foldings \cite{Hwang2015}.

If this scenario is working, then effectively there are two stages of inflation, where (1) the universe started from a massive potential $V_{0} (1 + \mu_{2}^{2} \phi_{2}^{2}/2)$ with $\mu_{2} < \sqrt{6\pi}$ and (2) later the universe experiences the inflation along the $V_{1}(\phi_{1})$ direction. This two-step inflation will give observational consequences. Originally there are two field directions, but for simplicity, we can assume that the field dynamics is gentle enough and this process will be well approximated by a single field model, where it initially started from a massive potential and eventually follows the usual inflation scenario, e.g., the Starobinsky model \cite{Starobinsky1980}. Therefore, finally, the following model is very good to demonstrate this scenario as well as its possible observational consequences:
\begin{equation}
	V=V_0 \left[ \left(1-e^{-\sqrt{\frac{2\kappa^2}{3}}\phi}\right)^2+\frac{1}{2}\mu^2(\phi-\phi_0)^2 \left( \frac{1}{2}+\frac{1}{\pi}\arctan{\frac{\phi-\phi_0}{\Delta}} \right) \right].
\end{equation}

\section*{Appendix B: Harmonic functions and numerical issue}

There are two reasons that CMB power spectrum $C^{TT}_\ell$ cannot be directly related to primordial power spectrum $\Delta^2_\mathcal{R}(n)$. First, the former is expanded in terms of the harmonic functions on the $S^3$ sphere, while the latter is expanded in the $S^2$ sphere. Therefore, what we really do is to project the $3$-dimensional spherical harmonics onto $2$-dimensional spherical harmonics. Secondly, after the reheating, the universe will filled with a mixture of different components, e.g., photons, neutrinos, baryons, cold dark matter, etc. We need to solve the Boltzmann equation along the the history of universe containing different component which makes the calculation difficult. It is unreliable to analytically solve the equations since there are several effects that need to be considered; however, there is a numerical code which helps us to solve the spectrum. In this appendix, we address the detailed relations between these two spectrum and the numerical method that we used.

Note that the CMB photon temperature anisotropy can be expanded by $S^2$ spherical harmonics:
\begin{equation} \label{eq:aperp} 
	\Delta_T=\frac{\delta{T}(\hat{n})}{T}=\sum_{\ell m}a_{\ell m}Y_{\ell m}(\hat{n}),
\end{equation}
where $\hat{n}$ is the radial direction pointing to the CMB, $Y_{\ell m}(\hat{n})$ is the usual spherical harmonics function, and $a_{\ell m}$ is the coefficient of the basis. The CMB temperature anisotropy is characterized by the correlation function $C^{TT}_\ell$, which is defined by
\begin{equation}
	C^{TT}_\ell = \frac{1}{2\ell+1}\sum_m\langle{a^*_{\ell m}a_{\ell m}}\rangle.
\end{equation}
We often use the scale invariant function $C_\ell$ to plot figures instead of $C^{TT}_\ell$, which is
\begin{equation}
	C_\ell \equiv \frac{\ell(\ell+1)}{2\pi}C^{TT}_\ell.
\end{equation}

On the other hand, the eigenvalues and eigenfunctions on $S^3$ sphere for the Laplace-Beltrami operator can be found as follows \cite{Niazy:2017jpn}:
%\footnote{Note that the mode $n=1$ and $n=2$ correspond to the pure gauge mode \cite{abbott1986general}, while we should not consider these two terms.}
\begin{eqnarray} 
	\nabla^2Q^k &=& -k^2_nQ^k, \\
	Q^k&=& Z_{n\ell m}\left(\chi,\theta,\varphi\right)=\Pi_{n\ell}(\chi)Y_{\ell m}\left(\theta,\varphi\right), \\
k^2_n &=& \frac{n^2-1}{R_c^2}, \quad \quad \quad n=1,2,...,
\end{eqnarray}
where $R_c$ is the curvature scale defined by $R_c=|K|^{-\frac{1}{2}}=|\Omega_k|^{-\frac{1}{2}}/aH$ and $\Pi_{n\ell}(\chi)$ is the hyperspherical Bessel function. %We can write the mode coefficients in terms of an $\ell$ dependent transfer function such as
%\begin{equation}
%	I_{\ell}=T_\ell(k)\Phi_k,
%\end{equation}
%where $\Phi_k$ is a random variable with
%\begin{equation}
%	\langle\Phi_k\Phi_{k'}\rangle=\Delta^2_\mathcal{R}(k)\delta_{kk'}.
%\end{equation}

Using these mode functions, one can simplify the power spectrum \cite{Lewis:2001}:
\begin{equation}
	C^{TT}_\ell = \frac{1}{16}\frac{1}{2\ell+1}\sum_k\Delta^2_\mathcal{R}(k_n)\left|T_\ell(k_n)\right|^2,
\end{equation}
where $T_{\ell}$ is a transfer function. After summing over $2\ell+1$ number of $m$-modes, $\sum_k=\sum_m\int dk/k$, the spectrum for the flat universe is
\begin{equation}
	C^{TT}_\ell = \frac{1}{16}\int\frac{dk_n}{k_n}\Delta^2_\mathcal{R}(k_n)\left|T_\ell(k_n)\right|^2,
\end{equation}
while for a closed universe, the integral is replaced by the discrete sum %\footnote{However, if the universe is nearly flat, the discrete wavenumbers that we have to sum is large, and hence we can sample the wavenumbers similar to that of the flat model; see chapter 7 of \cite{Lewis:2001}.}
\begin{equation}
	\int \frac{dk}{k}\rightarrow\sum_n\frac{n}{n^2-1}.
\end{equation}
Finally, we obtain
\begin{equation}
	C^{TT}_\ell=\frac{1}{16}\sum_n\frac{n}{n^2-1}\Delta^2_\mathcal{R}(n) \left|T_\ell(n)\right|^2.
\end{equation}
Here, the transfer function $T_\ell(n)$ can be obtained using the integral solution
\begin{equation}
	T_\ell(n)=\int^{t_0}_{0}dt\; \Pi_{n\ell}(\chi)S(n,t),
\end{equation}
where $t$ represents the traveling time of the photon along the geodesics, $t_0$ is the present observer's time, $\Pi_{n\ell}$ is the radial hyperspherical Bessel function, and $S(n,t)$ is a sum of the different terms that can generate temperature anisotropies including the gravitational redshift, the Doppler shift, and the intrinsic density fluctuation of the photon and baryon fluids, etc. The explicit form of $S(n,t)$ is in \cite{Zaldarriaga:1997va}. Since it contains complicated details with different factors, numerical techniques are required.

\newpage

%%%%%%%%%%%%%%%%%%%%%%%%%%%%%%%%%%%%%%%%%%%%%%%%%%%%%%%

\end{document}